\documentclass[fleqn,10pt]{wlscirep}
\usepackage{bm}
\usepackage{amssymb}
\usepackage{graphicx}
\usepackage{amsmath}
\usepackage{textcomp}
\usepackage{multirow}

\title{Second order optical nonlinearity of graphene due to electric quadrupole and magnetic dipole effects}

\author[1,2,*]{J. L. Cheng}
\author[1]{N. Vermeulen}
\author[2]{J. E. Sipe}
\affil[1]{Brussels Photonics Team (B-PHOT), 
  Department of Applied Physics and Photonics (IR-TONA), 
  Vrije Universiteit Brussel, Pleinlaan 2, 1050 Brussel,
  Belgium}
\affil[2]{Department of Physics and Institute for Optical Sciences,
  University of Toronto, 60 St. George Street, Toronto, Ontario,
  Canada M5S 1A7}
\affil[*]{Correspondence to jinluocheng.phys@gmail.com}
\begin{abstract}
We present a practical scheme to separate the contributions of the
electric quadrupole-like and the magnetic dipole-like effects to the
forbidden second order optical nonlinear response of graphene, and
give analytic expressions for the second order optical conductivities,
calculated from the independent particle approximation, 
with relaxation described in a phenomenological way. We predict
strong second order nonlinear effects, including second harmonic generation, photon drag, and difference
frequency generation. We discuss in detail the controllability of these effects by tuning the chemical potential, taking advantage of the dominant role played by interband optical transitions in the response.
\end{abstract}

\begin{document}

\flushbottom
\maketitle

\thispagestyle{empty}

\section*{Introduction}
Graphene is being enthusiastically
explored for potential applications in plasmonics, optoelectronics,
and photonics \cite{Nat.Photon._4_611_2010_Bonaccorso}, due to its
unique optical properties. They arise from the linear
dispersion of gapless Dirac fermions as
well as the ability to tune the Fermi energy with relative ease, by
either chemical doping \cite{J.Mater.Chem._21_3335_2011_Liu} or
applying a gate voltage \cite{Science_320_206_2008_Wang, Sci.Rep._4_6559_2014_Zhang}. With
the large optical nonlinearity predicted theoretically
\cite{Europhys.Lett._79_27002_2007_Mikhailov,NewJ.Phys._16_53014_2014_Cheng,Phys.Rev.B_91_235320_2015_Cheng,Phys.Rev.B_93_85403_2016_Mikhailov} and observed experimentally \cite{Phys.Rep._535_101_2014_Glazov},
graphene is also a potential resource of optical nonlinear functionality for photonic devices, including saturable
absorbers, fast and compact electo-optic modulators, and optical switches. Taking into account the maturing integration of
graphene onto silicon-based chips, the utilization of the
optical nonlinearity of graphene opens up new opportunities for the
realization of nonlinear integrated photonic circuits.

Due to the inversion symmetry of the graphene crystal, the first
nonvanishing nonlinear effect is the third order nonlinearity. In
spite of the one-atom thickness of graphene, strong third order  nonlinear
effects have
been demonstrated
\cite{NewJ.Phys._16_53014_2014_Cheng,Phys.Rev.B_91_235320_2015_Cheng} 
including parametric frequency conversion, third
harmonic generation, Kerr effects and two photon
absorption, and two color coherent current injection. The extracted effective nonlinear
coefficients are incredibly large, with values orders of magnitude
larger than those of usual semiconductors or metals. When fundamental 
photon frequencies $\omega_i$ are much smaller than the chemical
potential, as occurs in THz experiments on doped graphene,  the nonlinear
optical response is dominated by the intraband transitions
\cite{Europhys.Lett._79_27002_2007_Mikhailov,Phys.Rev.B_91_235320_2015_Cheng},
occurring 
mostly around the Fermi surface, and the third order optical
conductivities  have a typical frequency dependence\cite{Europhys.Lett._79_27002_2007_Mikhailov,Phys.Rev.B_91_235320_2015_Cheng} of $\propto
(\omega_1\omega_2\omega_3)^{-1}$ in the absence of relaxation. 
For photon energies in the near infrared to visible, the nonlinear
processes are dominated by the interband transitions and the mixing of
interband and intraband transitions \cite{Phys.Rev.B_91_235320_2015_Cheng}. 
In the presence of an energy gap induced by a suitable chemical potential \cite{NewJ.Phys._16_53014_2014_Cheng}, which
behaves as an energy gap in semiconductors, novel features arise in
the nonlinear optical response that cannot be easily found in semiconductors
or metals. These nonlinearities are both large and tunable, and promise
a new functionality in the design of the nonlinear optical properties
of integrated structures. The theoretical results based on the
independent particle approximation predict third order optical nonlinearities of graphene orders of magnitude smaller
than the experimental values
\cite{NewJ.Phys._16_53014_2014_Cheng,Phys.Rev.B_91_235320_2015_Cheng},
and the reason for the discrepancy has not been identified. 

The second order nonlinear optical response of graphene is forbidden
in the usual dipole approximation. However, it can arise due to a
number of effects \cite{Phys.Rep._535_101_2014_Glazov,
  Opt.Express_22_15868_2014_Cheng,Phys.Rev.B_91_235320_2015_Cheng}: (1) When
the variation of the electromagnetic field over the graphene is taken
into account, contributions analogous to those due to magnetic dipole
and electric quadrupole effects in centrosymmetric atoms or molecules
arise \cite{Europhys.Lett._79_27002_2007_Mikhailov,J.Phys.Condens.Matter_20_384204_Mikhailov,JETPLett._93_366_2011_Glazov,Phys.Rev.B_84_45432_2011_Mikhailov}; (2) at an asymmetric interface between graphene
and the substrate the centre-of-inversion symmetry is broken, and
second order nonlinearities are allowed \cite{Appl.Phys.Lett._95_261910_2009_Dean,Phys.Rev.B_82_125411_2010_Dean,Phys.Rev.B_85_121413_2012_Bykov,NanoLett._13_2104_2013_An,Phys.Rev.B_89_115310_2014_An}; (3) similarly, 
the symmetry can be locally broken due to natural curvature fluctuations
of suspended graphene \cite{Appl.Phys.Lett._105_151605_2014_Lin}; (4) the application of a dc electric
field can be used to generate an asymmetric steady state, and a second
order nonlinear optical response can then arise through the third
order nonlinearity
\cite{NanoLett._12_2032_2012_Wu,Phys.Rev.B_85_121413_2012_Bykov,J.Nanophoton._6_61702_2012_Avetissian,NanoLett._13_2104_2013_An,Phys.Rev.B_89_115310_2014_An,Opt.Express_22_15868_2014_Cheng}. 
 Second order optical responses due to the first effect
have been shown to be important for photon drag (dynamic Hall effects) \cite{Ganichev_2010,Phys.Rev.B_81_165441_2010_Entin,Phys.Rep._535_101_2014_Glazov}, second
harmonic generation (SHG) \cite{Phys.Rev.B_84_45432_2011_Mikhailov}, and difference frequency generation
(DFG)
\cite{Phys.Rev.Lett._112_55501_2014_Yao,Nat.Phys._12_124_2015_Constant,Phys.Rev.B_93_235422_2016_Tokman}.
[{\it Note added.} After the submission, we became aware of related works in preprints
  \cite{Phys.Rev.B_94_195442_2016_Wang,arXiv:1610.04854}. Overlapping 
  results in these papers are in agreement in the absence of
  relaxation time.] However, in most of the studies of these phenomena 
only intraband transitions were considered \cite{Ganichev_2010,Phys.Rev.B_84_45432_2011_Mikhailov,Phys.Rev.Lett._112_55501_2014_Yao}.  As with the third order
optical response, the contribution of interband transitions to the second
order nonlinearity can lead to a rich and tunable nonlinear optical response \cite{Phys.Rev.B_81_165441_2010_Entin,Phys.Rev.B_93_235422_2016_Tokman}. In this work we present analytic results for the second
order conductivities of graphene induced by the electric quadrupole-like and magnetic dipole-like effects, 
within the independent particle approximation and with relaxation
processes described phenomenologically.

\section*{Results}
\subsection*{Model}
We consider the charge current response of a graphene monolayer to electromagnetic fields $\bm
E(\bm r,t;z)$ and $\bm B(\bm r,t;z)$ with $\bm r=x\hat{\bm
  x}+y\hat{\bm y}$, and focus on the second order
conductivity $\sigma^{(2);dab}(\bm q_1,\omega_1;\bm q_2,\omega_2)$,
which is defined perturbatively for the weak fields from the second order current response
\begin{equation}
  J^{(2);d}(\bm r, t;z) = \delta(z) \int\frac{d\bm q_1d\bm q_2}{(2\pi)^4}\int
  \frac{d\omega_1 d\omega_2}{(2\pi)^2} e^{-i(\omega_1+\omega_2)t}
  e^{i(\bm q_1+\bm q_2)\cdot\bm r} \sigma^{(2);dab}(\bm
  q_1,\omega_1;\bm q_2,\omega_2) E^a_{\bm q_1\omega_1}(z)E^b_{\bm q_2\omega_2}(z)\,.\label{eq:defsigma2}
\end{equation}
Here the graphene layer is put at $z=0$, $E^a_{\bm q\omega}(z)$ is the in-plane Fourier transformation of the
electric field 
\begin{equation}
E^a(\bm r,t;z) = \int\frac{d\bm qd\omega}{(2\pi)^3}
e^{i\bm q\cdot\bm r-i\omega t} E^a_{\bm q\omega}(z)\,,\label{eq:ftef}
\end{equation}
the Roman superscript letters stand for the Cartesian directions $x$ or $y$,
the repeated superscripts imply a sum over all in-plane components, and
$\sigma^{(2);dab}(\bm q_1,\omega_1;\bm q_2,\omega_2)$ can be taken to
be symmetric in its components and arguments, $\sigma^{(2);dab}(\bm
q_1,\omega_1;\bm q_2,\omega_2) = \sigma^{(2);dba}(\bm q_2,\omega_2;\bm
q_1,\omega_1)$, without loss of generality. 
In writing Eq.~(\ref{eq:defsigma2}) we neglect any
response of the graphene to electric fields in the $z$ direction,
in line with the usual models for excitation around the Dirac points; 
thus the Cartesian components only range over $x$ and $y$. There
is no term involving the magnetic field $\bm B(\bm r,t;z)$
in Eq.~(\ref{eq:defsigma2}), but it is not neglected. Below we sketch the outline of
the derivation from the minimal coupling Hamiltonian, involving
the vector and scalar potentials. Keeping powers of $\bm q$ 
in the expansion of the vector potential introduces the magnetic field,
but the final result can be written in the form of Eq.~(\ref{eq:defsigma2}) in agreement
with the usual convention in nonlinear optics. However, since we focus
on the response at reasonably long wavelengths,  {\it i.e.}, $\hbar v_F |\bm q_i|\ll
|\hbar\omega_i|$ and $\hbar v_F |\bm q_1+\bm q_2|\ll |\hbar
\omega_1+\hbar\omega_2|$ with the electron Fermi velocity $v_F$, then the conductivity can be
expanded as 
\begin{equation}
  \sigma^{(2);dab}(\bm
  q_1,\omega_1;\bm q_2,\omega_2) \approx S^{dabc}(\omega_1,\omega_2)q_1^c +
  S^{dbac}(\omega_2,\omega_1) q_2^c\,.
\end{equation}
We have used the zero second order response to uniform fields
$\sigma^{(2);dab}(\bm 0,\omega_1;\bm 0,\omega_2)=0$, due to the
inversion symmetry of the graphene crystal structure. For the 
$D_{6h}$ crystal symmetry of graphene, fourth-order tensors $S^{dabc}$ have only
three independent in-plane components $S^{xxyy}$, $S^{xyyx}$, and
$S^{xyxy}$, and in total eight nonzero in-plane 
components $S^{yyxx}=S^{xxyy}$, $S^{yxxy}=S^{xyyx}$, $S^{yxyx}=S^{xyxy}$, and
$S^{yyyy}=S^{xxxx}= S^{xxyy} + S^{xyxy} + S^{xyyx}$. 

The response coefficients $S^{dabc}(\omega_{1},\omega_{2})$
completely characterize the second order optical response in the small
$|\bm q|$ limit. To calculate them, we begin by writing the minimal coupling
Hamiltonian as $\hat{H}=\hat{H}_0 -\frac{1}{2} e
[\bm A(\bm r,t;z)\cdot \hat{\bm v} + \hat{\bm v}\cdot\bm A(\bm r,t;z)]
+ e\phi(\bm r,t;z)$, where $e=-|e|$ is the electron charge, $\hat{H}_0$ is the unperturbed Hamiltonian,
$\hat{\bm v}$ is the velocity operator in the absence of an external field, $\bm
A(\bm r,t;z)$ and $\phi(\bm r,t;z)$ are the vector and scalar
potentials, respectively. 
Due to the linear dispersion relation of graphene around the Dirac
points, any higher order terms in $\bm A(\bm r,t;z)$ can
be neglected, unlike the situation in usual semiconductors, where
the calculation can be more difficult; the Zeeman interaction can
also be ignored. The vector and scalar potentials are then Fourier
expanded, as in Eq.~(\ref{eq:ftef}), and we write ${\cal A}_{\bm
  q\omega}^{\alpha}(z)=\phi_{\bm q\omega}(z)$
for $\alpha=0$ and ${\cal A}_{\bm q\omega}^{\alpha}(z)=A_{\bm q\omega}^{a}(z)$
for $\alpha=a=x,y,z$. The response current is then a functional of
${\cal A}^\alpha_{\bm q\omega}(z)$, and the formal second order perturbation
expansion gives
\begin{equation}
  J^{(2);d}(\bm r, t;z) = \delta(z) \int\frac{d\bm q_1d\bm q_2}{(2\pi)^4}\int
  \frac{d\omega_1 d\omega_2}{(2\pi)^2} e^{-i(\omega_1+\omega_2)t}
  e^{i(\bm q_1+\bm q_2)\cdot\bm r} W^{(2);d\alpha\beta}(\bm
  q_1,\omega_1;\bm q_2,\omega_2) {\cal A}^\alpha_{\bm
    q_1\omega_1}(z){\cal A}^\beta_{\bm q_2\omega_2}(z)\,.\label{eq:defw2}
\end{equation}
where repeated Greek indices range over $0,x,y,z$ . Not all components
of $W^{(2);d\alpha\beta}(\bm q_{1},\omega_{1};\bm q_{2},\omega_{2})$
are independent; they satisfy the Ward identity [see method], which is
associated with the invariance of the optical response to 
the choice of gauge. 

Consider first an electric field described by a scalar potential; then
we would have $W^{(2);d00}=(-i)^{2}q_{1}^{a}q_{2}^{b}\sigma^{(2);dab}$
in Eq.~(\ref{eq:md1}) [see method],
and by expanding both sides at small $\bm q_1$ and $\bm q_{2}$ we find 
\begin{equation}
    S^{dabc}(\omega_1,\omega_2) + S^{dcba}(\omega_1,\omega_2)  = -\left.\frac{\partial^3W^{(2);d00}(\bm q_1,\omega_1;\bm
      q_2,\omega_2)}{\partial q_1^a\partial q_1^c\partial
      q_2^b}\right|_{\bm q_1=\bm q_2=0}\,.\label{eq:sq}
\end{equation}
We define $S_Q^{dabc}(\omega_1,\omega_2) \equiv \left[S^{dabc}(\omega_1,\omega_2) + S^{dcba}(\omega_1,\omega_2)\right]/2$. 
For graphene, Eq.~(\ref{eq:sq}) can be used to determine the values of $S_Q^{xyxy}=S^{xyxy}$, $S_Q^{xxyy}=(S^{xxyy}+S^{xyyx})/2$, and further
$S_Q^{xxxx}=2S_Q^{xxyy}+S_Q^{xyxy}=S^{xxxx}$. However, the individual terms of
$S^{xxyy}$ and $S^{xyyx}$ cannot be obtained from
$W^{(2);d00}$. In general we can write $S^{dabc}(\omega_{1},\omega_{2})\equiv S_{Q}^{dabc}(\omega_{1},\omega_{2})+S_{M}^{dabc}(\omega_{1},\omega_{2})$ with defining $S_M^{dabc}(\omega_1,\omega_2) \equiv  \left[S^{dabc}(\omega_1,\omega_2) -
S^{dcba}(\omega_1,\omega_2)\right]/2$. Considering an electric field described by a vector
potential, we get $\sigma^{(2);dab}=W^{(2);dab}/(i^2\omega_1\omega_2)$ in
Eq.~(\ref{eq:md2}) [see method], and by expanding both sides at small $\bm q_1$ and $\bm q_2$
we find 
\begin{equation}
S_M^{dabc}(\omega_1,\omega_2)  = \frac{1}{2i\omega_1 i \omega_2}
\left[\frac{\partial W^{(2);dab}(\bm q_1,\omega_1;\bm 0,\omega_2)}{\partial q_1^c}-\frac{\partial W^{(2);dcb}(\bm q_1,\omega_1;\bm 0,\omega_2)}{\partial q_1^a}\right]\,.
\end{equation}
Of course, we could have used Eq.~(\ref{eq:md2}) as an expression for
all components of $\sigma^{(2);dab}(\bm q_1,\omega_1;\bm q_2,\omega_2)$
directly, and then for $S^{dabc}_Q$; in the relaxation free limit, we
have checked that the results of $S^{dabc}_Q$ calculated using the vector
potential are the same as those using the scalar potential only. The simple relaxation time approximation
used here is $not$ gauge invariant, which could be recovered by a postprocessing method \cite{Phys.Rev.B_1_2362_1970_Mermin}. In this work, the different calculations give differences only on the order of the relaxation parameters. We leave the gauge invariant relaxation time approximation for a future work.

For atoms, or molecules with center-of-inversion symmetry,
the kind of ``forbidden'' second order processes we are discussing
here can be identified with electric quadrupole and magnetic dipole
interactions, as opposed to the usual ``electric dipole interactions''
that typically govern the first order response. Here, however, in
a model where electrons are free to move through the graphene, there
is no simple way to clearly identify these two processes. We note
that while the expression for the full $S^{dabc}(\omega_{1},\omega_{2})$
can be derived solely from considering the vector potential, as mentioned
above, only its contribution $S_{Q}^{dabc}(\omega_{1},\omega_{2})$
can be identified by considering only the scalar potential. Since
quadrupole interactions in atoms and molecules exist if only scalar
potentials are introduced, while magnetic dipole interactions $require$
a vector potential for their description, we take this as a motivation
for ascribing quadrupole effects (or, more properly, quadrupole-like
effects) to $S_{Q}^{dabc}(\omega_{1},\omega_{2})$ , and for ascribing
magnetic dipole effects (or, more properly, magnetic-dipole-like effects)
to $S_{M}^{dabc}(\omega_{1},\omega_{2})$ . The independent nonzero
components of these tensors are $S_{Q}^{xxyy}(\omega_{1},\omega_{2})$,
$S_{Q}^{xyxy}(\omega_{1},\omega_{2})$, and $S_{M}^{xxyy}(\omega_{1},\omega_{2})$,
and in terms of them the second order current can be written as
\begin{align}
  \bm J^{(2)}(\bm r,t;z) =&2\delta(z) \int\frac{d\bm q_1 d\omega_1 d\bm
    q_2d\omega_2}{(2\pi)^6}e^{-i(\omega_1+\omega_2) t} e^{i(\bm q_1+\bm
  q_2)\cdot\bm r} \Big\{  S_M^{xxyy}(\omega_1,\omega_2)[\bm q_1\times \bm E_{\bm
    q_1\omega_1}(z)]\times\bm E_{\bm q_2\omega_2}(z)\notag\\
+S_{Q}^{xxyy}&(\omega_1,\omega_2) [\bm E_{\bm
  q_1\omega_1}(z) \bm q_1\cdot
 \bm E_{\bm
  q_2\omega_2}(z) +\bm q_1 \bm E_{\bm q_1\omega_1}(z) \cdot
 \bm E_{\bm
  q_2\omega_2}(z)]  +  S^{xyxy}_Q(\omega_1,\omega_2) \bm q_1\cdot \bm E_{\bm
    q_1\omega_1}(z) {\bm E}_{\bm q_2\omega_2}(z) 
\Big\}\,. \label{eq:jqm}
\end{align}
We now present a microscopic theory to calculate the tensor components.

We describe the low energy electronic states $\psi_{s\bm k}(\bm r;z)$ at
 band index $s=\pm$ and wave vector $\bm k$ by a widely used two-band
tight binding model based on  the carbon $2p_z$ orbitals \cite{NewJ.Phys._16_53014_2014_Cheng}. 
Ignoring all response to the $z$-component of the electric field, the total Hamiltonian can be written as  $H=H_0+H_{el}+H_{scat}$ with the
unperturbed Hamiltonian $H_0 = \sum_{s}\int_{\text{BZ}} d\bm k \varepsilon_{s\bm k}a_{s\bm k}^\dag(t)
  a_{s\bm k}(t)$ and
\begin{equation}
  H_{el}= \int\frac{d\bm q d\omega}{(2\pi)^3}
  e^{-i\omega t}  \sum_{s_1s_s2}\int_{\text{BZ}}d\bm k e{\cal A}_{\bm
    q,\omega}^\alpha(0){\cal W}^\alpha_{s_1\bm k+\bm q;s_2\bm k}a_{s_1\bm k+\bm q}^\dag(t)
  a_{s_2\bm k}(t)\,.
\end{equation}
Here $\varepsilon_{s\bm k}$ is the electron band energy, $a_{s\bm k}(t)$ is an annihilation
operator associated with $\psi_{s\bm k}(\bm r;z)$, the integration is
over one Brillouin zone (BZ), and $H_{scat}$ is the scattering
Hamiltonian described below phenomenologically. The interaction matrix elements
at $\alpha=0$ give ${\cal W}^0_{s_1\bm k_1;s_2\bm k_2}=D_{s_1\bm
  k_1;s_2\bm k_2}$, which is the matrix element of a plane wave $e^{i(\bm
  k_1-\bm k_2)\cdot\bm r}$ between states $s_1\bm k_1$ and $s_2\bm
k_2$;  the other three components are ${\cal W}^d_{s_1\bm k_1;s_2\bm
  k_2}=-{\cal V}^d_{s_1\bm k_1;s_2\bm k_2}$ where ${\cal V}^d_{s_1\bm k_1;s_2\bm k_2}=\sum_s( v_{s_1s\bm
  k_1}^d D_{s\bm k_1,s_2\bm k_2} + D_{s_1\bm k_1,s\bm k_2}v_{ss_2\bm
  k_2}^d)/2$ is the matrix element of the velocity density, with $\bm
v_{s_1s_2\bm k}$ being the velocity matrix elements in the absence of
an external field. The dynamics of the system is described by a single
particle density matrix $\rho_{s_1\bm k_1;s_2\bm k_2}(t)=\langle
a_{s_2\bm k_2}^{\dag}(t) a_{s_1\bm k_1}(t)\rangle$, which satisfies the
equation of motion
\begin{align}
    i\hbar\frac{\partial \rho_{s_1\bm k_1;s_2\bm k_2}(t)}{\partial t}
    =& (\varepsilon_{s_1\bm k_1}-\varepsilon_{s_2\bm k_2})
    \rho_{s_1\bm k_1;s_2\bm k_2}(t)+ \int\frac{d\bm qd\omega}{(2\pi)^3}
e^{-i\omega t}e{\cal A}^\delta_{\bm q\omega}\sum_{s}\big[ {\cal W}^\delta_{s_1\bm k_1;s\bm k_1-\bm
    q} \rho_{s\bm k_1-\bm q;s_2\bm k_2}(t)  \notag\\
& - \rho_{s_1\bm k_1;s\bm
    k_2+\bm q}(t){\cal W}^\delta_{s\bm k_2+\bm q;s_2\bm k_2}\big]  - \Gamma [\rho_{s_1\bm k_1;s_2\bm k_2}(t) -
  \rho_{s_1\bm k_1,s_2\bm k_2}^0]\,. \label{eq:eom}
\end{align}
Here the last term describes the scattering effects phenomenologically with one relaxation energy $\Gamma$, and
$\rho_{s_1\bm k_1,s_2\bm k_2}^0=\delta_{s_1s_2}\delta(\bm k_1-\bm
k_2)n_{s_1\bm k}$ is the initial carrier distribution without any
external fields, where $n_{s_1\bm k}=F_\mu(\varepsilon_{s_1\bm k},T)$
and $F_\mu(x,T)=[1+e^{\beta(x-\mu)}]^{-1}$ with $\beta=1/(k_BT)$ is the Fermi-Dirac distribution
at the chemical potential $\mu$ and the temperature $T$. We focus on
the current response $\bm J(\bm r,t;z) = \delta(z) \int d\bm q J_{\bm
  q}(t)e^{i\bm q\cdot\bm r}$ with $J^d_{\bm q}(t) = e\sum_{s_1s_2}\int_{\text{BZ}}\frac{d\bm k}{(2\pi)^2}
  {\cal V}^d_{s_1\bm k;s_2\bm k+\bm q}\rho_{s_2\bm k+\bm q,s_1\bm
    k}(t)$. The perturbation results are
\begin{align}
  W^{(2);d\alpha\beta}(\bm q_1,\omega_1;\bm q_2,\omega_2) =&
  \frac{1}{2}\left[\widetilde{W}^{(2);d\alpha\beta}
  (\bm q_1,\omega_1;\bm q_2,\omega_2)+\widetilde{W}^{(2);d\alpha\beta}
  (\bm q_2,\omega_2;\bm q_1,\omega_1)\right]\,,\\
\widetilde{W}^{(2);d\alpha\beta}(\bm q_1,\omega_1;\bm
q_2,\omega_2)=&g_s\sum_{s}\int_{\text{BZ}}\frac{d\bm k}{(2\pi)^2} e^3
\widetilde{\cal P}^{(2);d\alpha\beta}_{ss\bm k}(\bm q_1,\bm
q_2;\hbar\omega_1+\hbar\omega_2+i\Gamma,\hbar\omega_2+i\Gamma)n_{s\bm k}\,.\label{eq:w2}
\end{align}
Here $g_s=2$ is the spin degeneracy. The term $\widetilde{\cal P}^{(2);d\alpha\beta}_{s_1s_2\bm k}$ is given by
\begin{align}
  \widetilde {\cal P}^{(2);d\alpha\beta}_{\bm k}(\bm q_1,\bm q_2;w_0,w_2) =&
 \overline{\cal V}^d_{\bm k,\bm k+\bm q_1+\bm q_2}(w_0) {\cal W}^\alpha_{\bm k+\bm
    q_1+\bm q_2,\bm k+\bm
    q_2}\overline{\cal W}^\beta_{\bm k+\bm q_2,\bm k}(w_2)-
  \overline{\cal W}^\beta_{\bm
    k,\bm k-\bm q_2}(w_2)\overline{\cal V}^{d}_{\bm k-\bm q_2,\bm
    k+\bm q_1}(w_0) {\cal W}^\alpha_{\bm k+\bm q_1,\bm k}\notag\\
   - {\cal W}^\alpha_{\bm
    k,\bm k-\bm q_1}&\overline{\cal V}^{d}_{\bm k-\bm q_1,\bm
    k+\bm q_2}(w_0) \overline{\cal W}^\beta_{\bm k+\bm q_2,\bm k}(w_2)
  + \overline{\cal W}^\beta_{\bm k,\bm k-\bm q_2}(w_2){\cal W}^\alpha_{\bm k-\bm q_2,\bm
    k-\bm q_1-\bm q_2}\overline{\cal V}^d_{\bm k-\bm q_1-\bm q_2,\bm k}(w_0)\,,\label{eq:2ndv}
\end{align}
where each quantity is expressed as a $2\times2$ matrix with
abbreviated band index, and 
\begin{equation}
  \overline{\cal W}^\delta_{s_1\bm k_1,s_2\bm k_2}(w_0) = \frac{{\cal
      W}^\delta_{s_1\bm k_1,s_2\bm k_2}}{w_0-\varepsilon_{s_1\bm
      k_1}+\varepsilon_{s_2\bm k_2}}\,,\quad   \overline{\cal V}^d_{s_1\bm k_1,s_2\bm k_2}(w_0) = \frac{{\cal
      V}^d_{s_1\bm k_1,s_2\bm k_2}}{w_0-\varepsilon_{s_2\bm
      k_2}+\varepsilon_{s_1\bm k_1}}\,.
\end{equation}

In the following we explicitly indicate the $\mu$ and $T$ dependence of
$S^{dabc}_{Q;\mu,T}$ and $S^{dabc}_{M;\mu,T}$. Based on the electron-hole symmetry in our tight binding
model, and the time and space inversion
symmetries of the graphene crystal, we find $\widetilde{\cal P}^{(2);d\alpha\beta}_{++\bm
  k}=\widetilde{\cal P}^{(2);d\alpha\beta}_{--\bm k}$, which indicates that the contributions of the electrons and holes to the second
order response coefficients are the same.  At zero temperature, when
the chemical potential $\mu\to+\infty$, all the states are filled and
there should be no response, $S_{Q/M;+\infty,0}^{dabc}(\omega_1,\omega_2) =
0$. Since the electrons and holes lead to the same contribution we have
$S_{Q/M;+\infty,0}^{dabc}(\omega_1,\omega_2) = 2
S_{Q/M;0,0}^{dabc}(\omega_1,\omega_2)$, and so we have $
S_{Q/M;0,0}^{dabc}(\omega_1,\omega_2) = 0$ as well. This is an important result, because in general $S_{Q/M;0,0}^{dabc}(\omega_1,\omega_2)$ cannot be directly evaluated if only the low energy electronic excitation is available. Utilizing the linear dependence of $n_{s\bm k}$ in Equation~(\ref{eq:w2}), the calculation of $S_{Q/M;\mu,0}^{dabc}-S_{Q/M;0,0}^{dabc}\equiv S_{Q/M;\mu,0}^{dabc}$ depends on the electronic states around the Dirac points only.

\subsection*{Conductivity in the linear dispersion approximation}
For visible or infrared light, the optical transitions occur mostly
around the Dirac point, where the linear dispersion
approximation is widely used. The two Dirac points are at $\bm K_1=\frac{1}{3}\bm b_1 + \frac{2}{3}\bm b_2$ and $\bm
K_2=\frac{2}{3}\bm b_1+\frac{1}{3}\bm b_2$, with the primitive reciprocal lattice vectors $\bm
b_1=\frac{2\pi}{a_0}(\frac{1}{\sqrt{3}}\hat{\bm x}-\hat{\bm y})$, $\bm
b_2=\frac{2\pi}{a_0}(\frac{1}{\sqrt{3}}\hat{\bm x}+\hat{\bm y})$, and
lattice constant $a_0=2.46$\AA.  Noting that for $\bm k$ around $\bm K_1$
we have $\bm k=\bm K_1+\bm\kappa$ with $\bm\kappa=\kappa(\cos\theta_{\bm\kappa}\hat{\bm
  x}+\sin\theta_{\bm\kappa}\hat{\bm y})$, in the linear dispersion
approximation we have $\varepsilon_{s\bm k} = s \hbar v_F \kappa$ and 
\begin{eqnarray}
 D_{s_1\bm k; s_2\bm k+\bm q} = \frac{1}{2} \left[ 1 + s_1s_2
    e^{i(\theta_{\bm\kappa} - \theta_{\bm \kappa+\bm
        q})}\right]\,,\quad   {\cal V}^x_{s_1\bm k;s_2\bm k+\bm q} =  \frac{1}{2}\hbar v_F (s_1 e^{i\theta_{\bm\kappa}} +
  s_2e^{-i\theta_{\bm \kappa+\bm q}})\,,\quad {\cal V}^y_{s_1\bm k;s_2\bm k+\bm q} =  \frac{1}{2}\hbar v_F (-is_1 e^{i\theta_{\bm\kappa}} +
  is_2e^{-i\theta_{\bm \kappa+\bm q}})
\end{eqnarray}
with the Fermi velocity $v_F =
{\sqrt{3}a_0\gamma_0}/{(2\hbar)}$ and the hopping parameter $\gamma_0=2.7$~eV. The appropriate expression around the
other Dirac point can be obtained using inversion
symmetry. We perform the integration first over the angle
$\theta_{\bm\kappa}$ and then  over $\kappa$. Utilizing
$S_{Q/M;0,0}^{dabc}(\omega_1,\omega_2) = 0$ we find the results 
\begin{align}
  S^{xxyy}_{Q;\mu,0}(\omega_1,\omega_2) =&
  \text{sgn}(\mu)\sigma_2\left[
    \frac{8\mu^2}{(w_0^2-4\mu^2)^2}\frac{-i\Gamma}{w_2(\hbar\omega_2)}+\frac{4\mu^2}{w_0^2-4\mu^2}\frac{w_0-2w_1}{w_1^2(\hbar\omega_2)^2}+\frac{4\mu^2}{w_1^2-4\mu^2}\frac{1}{w_0(\hbar\omega_2)^2}\right]\,,\notag\\
  S^{xyxy}_{Q;\mu,0}(\omega_1,\omega_2)
  =&-S^{xxyy}_{Q;\mu,0}(\omega_1,\omega_2) - \text{sgn}(\mu)\sigma_2
  \frac{8\mu^2}{w_0^2-4\mu^2}\frac{w_1^2+w_2(w_0+w_1)}{w_0^2w_1^2w_2}\notag\\
  S^{xxyy}_{M;\mu,0}(\omega_1,\omega_2) = &\text{sgn}(\mu)\sigma_2 \frac{1}{\hbar\omega_1(\hbar\omega_2)^2}\left(-\frac{8\mu^2}{w_0^2-4\mu^2}+\frac{8\mu^2}{w_1^2-4\mu^2}\frac{2w_0-w_1}{w_0}\right)
\label{eq:mainresult}
\end{align}
Here $\sigma_2=\sigma_0{|e|(\hbar
  v_F)^2}/{(2\pi)}$, $\sigma_0=e^2/(4\hbar)$,
$w_1=\hbar\omega_1+i\Gamma$, $w_2=\hbar\omega_2+i\Gamma$, and
$w_0=\hbar(\omega_1+\omega_2)+i\Gamma$. 
Simply taking the limit $\mu\to+\infty$ in
Eq.~(\ref{eq:mainresult}) does not recover the result 
$S_{Q/M;+\infty,0}^{dabc}(\omega_1,\omega_2)=0$. This is not surprising,
because such a limit involves the
contributions from all electrons in the ``$-$'' band, most of which can
not be described by the linear dispersion. Nonetheless, the
contributions to the second order response from electrons close to the
Dirac points are well described by Eqs.~(\ref{eq:2ndv}) and (\ref{eq:mainresult}). Combined with
the fact that the conditions $S_{Q/M;0,0}^{dabc}(\omega_1,\omega_2)=0$
are verified by using the
symmetries of the system, the expression in Eq.~(\ref{eq:mainresult}) can be
used to describe the response coefficient for optical transitions
occurring around the Dirac points.

At finite temperature, we follow the technique
used in our previous work\cite{Phys.Rev.B_91_235320_2015_Cheng} to calculate the conductivities as
\begin{equation}
S^{dabc}_{Q/M;\mu,T}(\omega_1,\omega_2) = \beta \int_0^\infty dx
\{F_\mu(x,T)[1-F_\mu(x,T)]-F_\mu(-x,T)[1-F_\mu(-x,T)]\}S^{dabc}_{Q/M;x,0}(\omega_1,\omega_2)\,,\label{eq:tem}
\end{equation}
As opposed to the results
of calculations of the third order conductivities\cite{Phys.Rev.B_91_235320_2015_Cheng} along these lines, here all terms appearing
in Eq.~(\ref{eq:mainresult}) are well behaved in the integration of Eq.~(\ref{eq:tem}). 

The main results of this section are given in Eqs.~(\ref{eq:jqm}),
(\ref{eq:mainresult}),  and (\ref{eq:tem}). Within the linear dispersion
approximation that we have assumed, the results are analytic, and any calculation can be performed
directly. In the following we discuss the divergences (poles) of
the analytic expressions at zero temperature, and then give a quantitative analysis for different second
order optical nonlinear phenomena including SHG, one color dc current generation (including current
injection effects, photon drag (or dynamic Hall
effect)), and DFG.

\subsection*{Features and limitations of the result}
We begin by considering some special limits of the response
following from Eqs.~(\ref{eq:jqm}),
(\ref{eq:mainresult}),  and (\ref{eq:tem}).
In the limit of a chemical potential much greater than any other energies
involved, {\it i.e.}, $|\mu|\gg \hbar\omega_i,\Gamma, k_BT$, the
dominant contribution to the response is expected to come from the
intraband transitions between states around the Fermi
surface. We can isolate this 
contribution by considering the limit $\lim\limits_{x\to0}S^{dabc}_{\mu,T}(x
\omega_1,x\omega_2;x\Gamma)$ and keeping only the leading term that
varies as $\propto x^{-3}$, and then
setting $x=1$. In this limit we find for the intraband contribution
\begin{align}
  S_{Q;\text{intra}}^{xxyy}(\omega_1,\omega_2) =& -\text{sgn}(\mu)
  \sigma_2 \frac{1}{w_1^2w_0} \,,\quad 
  S_{Q;\text{intra}}^{xyxy}(\omega_1,\omega_2) = \text{sgn}(\mu) \sigma_2 \frac{2w_1(w_1+w_2)
    + 3w_2w_0}{w_1^2w_2w_0^2}\,,\notag\\
  S_{M;\text{intra}}^{xxyy}(\omega_1,\omega_2)=&-\text{sgn}(\mu)\sigma_2\frac{2}{\hbar\omega_1\hbar\omega_2w_0}\,.\label{eq:intra}
\end{align}
Note that except for a sign the results in Eq.~(\ref{eq:intra}) are
independent of the chemical potential, which through
Eq.~(\ref{eq:tem}) leads also to an insensitivity to the temperature. 
There are
two kinds of divergences that appear in Eq.~(\ref{eq:intra}). The first involves
the $w_{i}$ in the denominator; the $w_{i}$ never vanish at real
frequencies, and the divergences that would arise were $\Gamma$ to
vanish can be said to be ameliorated by the phenomenological relaxation
introduced. The second involves the $\omega_{i}$ in the expression
for $S_{M;\text{intra}}^{xxyy}(\omega_{1},\omega_{2})$ , and are unameliorated.
Even in the presence of relaxation these lead to divergences as
$\omega_{1}$ or $\omega_{2}$ vanishes, and they appear in the term where the vector
potential was used in the calculation. In fact, if we evaluate the
terms $S_{Q;\text{intra}}^{xxyy}(\omega_{1},\omega_{2})$
and $S_{Q;\text{intra}}^{xyxy}(\omega_{1},\omega_{2})$ using the
vector potential by Eq.~(\ref{eq:md2}) instead of the scalar potential,
we also find that they acquire unameliorated divergences. We emphasize
that in the limit of no relaxation, where unameliorated divergences
appear everywhere, the result in Eqs.~(\ref{eq:jqm}),
(\ref{eq:mainresult}),  and (\ref{eq:tem}) is independent of the gauge
used in the calculations; our results agree with those obtained by Tokman {\it et al.} \cite{Phys.Rev.B_93_235422_2016_Tokman}. It is just that the commonly used phenomenological
model we have introduced for relaxation is too simple to respect this
gauge invariance. So the divergences in our expression for $S_{M;\text{intra}}^{xxyy}(\omega_{1},\omega_{2})$
should not be taken seriously; they are artifacts of the relaxation
model, and have a parallel in the same way that unameliorated divergences
can arise in the linear response of a metal if such a relaxation model
is used in conjunction with the use of a vector potential to describe
the electric field. We will turn to a more sophisticated treatment
of the relaxation in a future work; in this paper our focus will be
on features of the response where $\omega_{1}$ and $\omega_{2}$ are
greater than $\Gamma/\hbar$ from zero, and thus the lack of amelioration
of the vector potential divergences will not be crucial.

Generalizing now beyond just the intraband response, we note that in
the $(\bm q,\omega)$ dependence of the linear
conductivity\cite{NewJ.Phys._8_318_2006_Wunsch,Phys.Rev.B_75_205418_2007_Hwang}
$\sigma^{(1);da}(\bm q,\omega)$, there are divergences that arise in the absence of relaxation when
one of the resonant conditions $\hbar v_{F}|\bm q|\approx\hbar\omega$
and $\hbar v_{F}|\bm q|\approx |\hbar\omega-2|\mu||$ is
met. The first is associated with intraband transitions, and the second
with interband transitions. Similar divergences arise here, some involving
combinations of wave vectors and frequencies, and their appearance
is evident in the denominator of Eq.~(\ref{eq:2ndv}). The general
resonant conditions could be met in photonic
structures, where $|\bm q|$ can be much larger than
$\omega/c$. For light incident from
vacuum where the magnitude of the incident wave vectors $|\bm
q_{j}|\leq \omega_{j}/c$, since $v_{F}\ll c$ the resonant conditions
for incident fields become $\omega_{j}=0$
for intraband transitions and $\omega_{j}=\pm2|\mu|/\hbar$
for interband transitions, as shown in the analytic expression for
$S_{\mu,0}^{dabc}(\omega_{1},\omega_{2})$. For the generated field at 
$\omega_1+\omega_2$ and $\bm q_1+\bm q_2$, the general resonant
condition may be satisfied due to the arbitrary choice of incident angle
\cite{Phys.Rev.Lett._112_55501_2014_Yao,Nat.Phys._12_124_2015_Constant}. Since the response to
the intraband transitions in Eq.~(\ref{eq:intra}) is weakly dependent on the chemical potential,
one must rely on the interband contribution to tune the resonant second
order response in graphene. All coefficients
$S^{dabc}_{\mu,T}(\omega_2,\omega_3)$ are odd functions of the
chemical potential, and at least proportional to
$\text{sgn}(\mu)\mu^2$. Among these, $S^{(2);xxxx}_{\mu,T}(\omega_2,\omega_3)\propto
\text{sgn}(\mu)\mu^4$. The form of the divergences indicates the temperature can
strongly affect the values of $S^{dabc}$ around these divergences.  At
room temperature, all these fine structures are greatly smeared out
even without the inclusion of the relaxation.

\subsection*{Second harmonic generation}
For a single plane wave of fundamental light incident on the graphene
sheet, which at $z = 0$ will give a field of the form
\begin{equation}
  \bm E(\bm r, t;0) = \bm E_0 e^{i\bm q_0\cdot\bm r-i\omega_0t} +
  c.c.\,, \label{eq:Esgmode}
\end{equation}
it is convenient to separate the components of the field parallel
and perpendicular to $\bm q_0$ as $E_0^\parallel = \hat{\bm
  q}_0\cdot\bm E_0$ and $E_0^\perp =
  \hat{\bm z}\cdot(\hat{\bm q}_0\times \bm
  E_0)$, respectively. In the notation of Eq.~(\ref{eq:ftef}) we have $\bm E_{\bm q\omega}(0) = (2\pi)^3 \delta(\bm q-\bm
  q_0)\delta(\omega-\omega_0) \bm E_0$. The generated second harmonic current is 
\begin{equation}
  \bm J_{\text{SHG}}(\bm r,t;z)  = \delta(z) e^{2i\bm q_0\cdot\bm r-2i\omega_0t}
  \Big\{\bm q_0 S_1(\omega_0) (E_0^\parallel)^2 + \bm q_0
  S_2(\omega_0) (E_0^\perp)^2  +\hat{\bm z} \times {\bm q_0} [S_1(\omega_0)-S_2(\omega_0)] E_0^\perp E_0^\parallel  \Big\} + c.c.\,,\label{eq:jshg}
\end{equation}
where 
\begin{equation}
S_1(\omega_0) = 2 S^{xxxx}_{\mu,T}(\omega_0,\omega_0)\,,\quad S_2(\omega_0) =
2S^{xyyx}_{\mu,T}(\omega_0,\omega_0)\,.
\end{equation}
Note that a current perpendicular to $\bm q_{0}$ arises only
when components of the electric field both parallel and perpendicular
to $\bm q_{0}$ are present, while a current in the direction
of $\bm q_0$ arises quite generally.

\begin{figure}[htp]
  \centering
  \includegraphics[width=15.9cm]{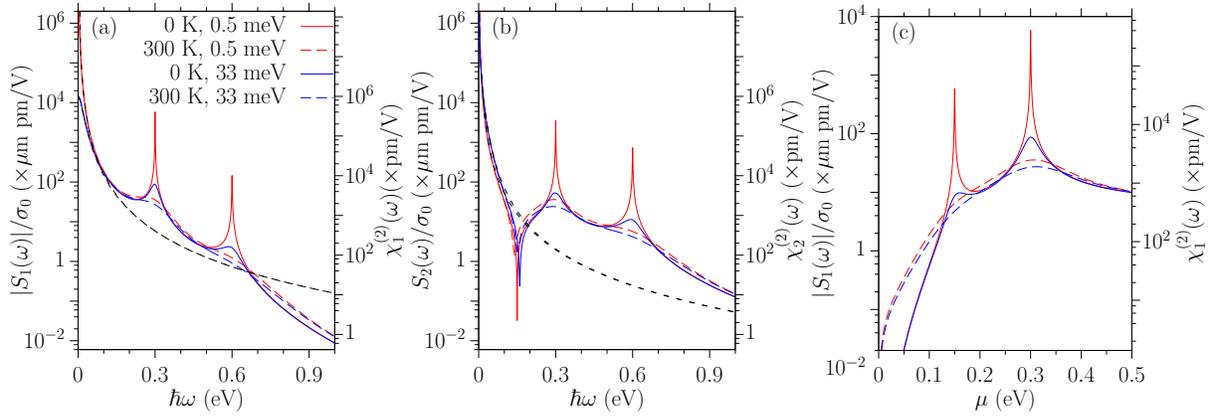}
  \caption{The response coefficients (a) $|S_1(\omega)|$ and (b)
    $|S_2(\omega)|$  for relaxation parameters $\Gamma=0.5$~meV
    and  $33$~meV  at the temperatures $T=0$ and $300$~K and chemical
    potential $\mu=0.3$~eV. The black dashed curves are the intraband
    contributions from Eq.~(\ref{eq:intra}). (c) shows the chemical
    potential dependence of $|S_1(\omega)|$ at $\hbar\omega=0.3$~eV
    for the same relaxation parameters and temperatures. The right
    $y$-axis shows the second order susceptibility for light that
    propagates parallel to the graphene sheet. }
  \label{fig:shg}
\end{figure}
In Fig.~\ref{fig:shg} (a) and (b) we show the response coefficients $S_1(\omega)$
and $S_2(\omega)$ for relaxation parameters $\Gamma=0.5$ and
$33$~meV at temperatures $T=0$ and $300$~K, for the chemical
potential $\mu=0.3$~eV. As $\omega\rightarrow0$ the description of relaxation is not valid,
as we discussed above, so we focus on the behavior away from
$\omega=0$. For zero temperature and a small relaxation parameter the
coefficient $S_{1}$ exhibits two peaks, one at $\hbar\omega=|\mu|$
and the other at $\hbar\omega=2|\mu|$; they follow from the analytic
expression in Eq.~(\ref{eq:mainresult}). As discussed above, the 
peaks are as expected for interband resonances, the first associated
with a two-photon resonance and the second with a one-photon resonance.
With an increase in the relaxation parameter or in the temperature
both peaks are lowered and broadened. Although the thermal energy at
$300$~K ($25.8$~meV) is slightly smaller than the relaxation parameter
$33$~meV, it affects both peaks more effectively, which follows from the form of the dependence on the chemical potential
in Eqs.~(\ref{eq:mainresult}) and (\ref{eq:tem}). The intraband contributions from
Eq.~(\ref{eq:intra}) are plotted as black dashed curves, which 
fit with the fully calculated results very well for photon energies
$\hbar\omega<0.1$~eV for the chosen parameters. 
The coefficient $S_2$ exhibits similar peaks at $\hbar\omega=|\mu|$
and $\hbar\omega=2|\mu|$, but there is also a dip at around
$\hbar\omega=|\mu|/2$. This  is also apparent from
Eq.~(\ref{eq:mainresult}), and is due to a cancellation of
contributions from $S_{Q}^{xxyy}$ and $S_{M}^{xxyy}$; in fact, at zero
temperature, $S_2(\mu/2)\propto\Gamma$ and so it would vanish
in the limit of no relaxation. 

For the extreme
case of $q_{0}=\omega_{0}/c$, corresponding to light propagating
parallel to the graphene sheet, it is natural to introduce an effective
SHG susceptibility. Identifying a nominal thickness $d_{gr}=3.3\textrm{\AA}$
for the graphene sheet, the effective susceptibility can be taken
as  $\chi_j^{(2)}(\omega_0) = {S_j(\omega_0)q_0}/{(-i\omega_0\epsilon_0 d_{\text{gr}})} =
{iS_j(\omega_0)}/{(c\epsilon_0 d_{\text{gr}})}$ with $j=1,2$; note that $\chi_{j}^{(2)}(\omega_{0})$
is simply proportional to $S_{j}(\omega_{0})$, but its introduction
makes it easy to compare the strength of the second order response
of graphene with that of materials with an allowed second-order
response. The values  of the $\chi_{j}^{(2)}(\omega_{0})$  are shown
on the right $y$-axis of Fig.~\ref{fig:shg}. At $\hbar\omega_0\sim
0.6$~eV, the values for $|\chi_1^{(2)}(\omega_0)|$ vary from
$10^5$~pm/V at $T=0,\Gamma=0.5$~meV, to $84$~pm/V at
$T=300,\Gamma=0.5$~meV, $154$~pm/V at $T=0,\Gamma=33$~meV,
and $63$~pm/V at $T=300,\Gamma=33$~meV. The temperature and
relaxation greatly reduce its magnitude. The contribution from
the intraband transition is about $51$~pm/V, which is insensitive to the
temperature and relaxation parameters at this photon energy. At optical frequencies
the values obtained here are orders of magnitude smaller than those
predicted for the current-induced SHG
\cite{Opt.Express_22_15868_2014_Cheng} of graphene and
the SHG \cite{Phys.Rev.B_92_235307_2015_Cheng} in a gapped graphene. This is
not surprising; effects dependent on the finite size of the wave vector
of light are typically weak. However, the values we find are still
larger than for most SHG
materials \cite{NanoLett._12_2032_2012_Wu}, where the process is
allowed, which indicates the strong second order optical response of
graphene, despite the fact that it must rely on the small wave vector
of light. 

In Fig.~\ref{fig:shg}(c) we show the chemical potential dependence of
$|S_1(\omega)|$ for a fixed photon energy $\hbar\omega=0.3$~eV. We see
that control of the
chemical potential can be used to change the size of the SHG coefficients,
especially at low temperature. 

The second order polarizability constitutes part
of the second order response, and has been investigated by Mikhailov 
\cite{Phys.Rev.B_84_45432_2011_Mikhailov}. The connection between the 
nonlinear conductivity discussed here and that polarizability follows
from the continuity equation $\partial_t n(\bm r, t;z)+
\bm\nabla_{\bm r}\cdot \bm J(\bm r,t;z) = 0$.  For the field in
Eq.~(\ref{eq:Esgmode}), the induced second order charge density is
identified as $
  n^{(2)}(\bm r,t;z) = \delta(z) e^{2i\bm q_0\cdot\bm
    r-2i\omega_0t}\alpha^{(2)}(\bm q_0,\omega_0)\phi_{\bm
    q_0\omega_0}^2(z)+c.c.$, 
from which we find $
   \alpha^{(2)}(\bm q_0,\omega_0) =-\frac{ 2q_0^d}{\omega_0}
   S^{dabc}(\omega_0,\omega_0)q_0^aq_0^bq_0^c=-\frac{q_0^4}{\omega_0}S_1(\omega_0)$. 
The intraband contribution without the inclusion of relaxation gives from
Eq.~(\ref{eq:intra}) as $\alpha^{(2)}(\bm q_0,\omega_0) = -\frac{3|e|^3
  q_0^4(\hbar v_F)^2}{8\pi(\hbar\omega_0)^4}$, which is in agreement
with Mikhailov's calculation \cite{Phys.Rev.B_84_45432_2011_Mikhailov}. This also confirms that his
expression contains only intraband contributions, and
as expected there is no contribution to the second order polarizability
from magnetic-dipole-like terms.

\subsection*{Photon drag and one color current injection}
\begin{figure}[!htpb]
  \centering
  \includegraphics[width=15.9cm]{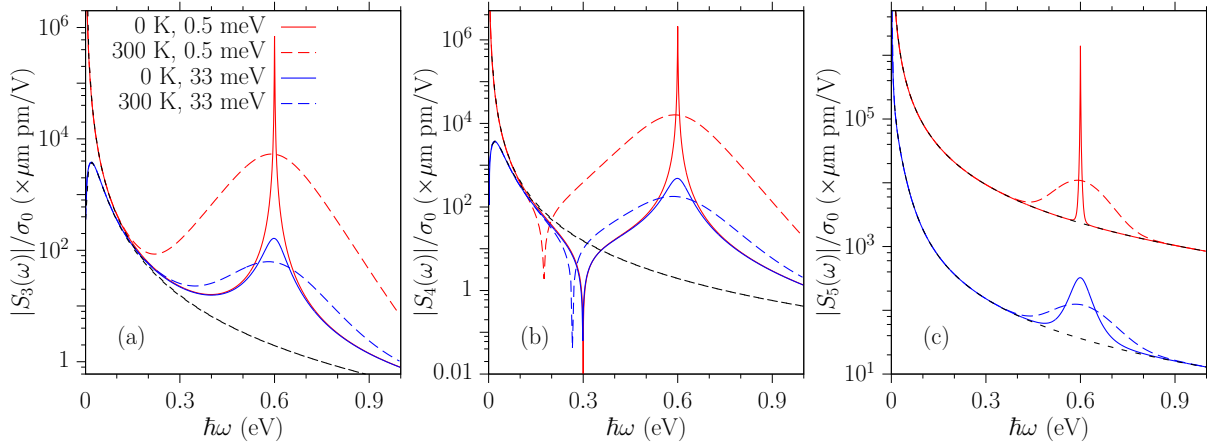}  
  \caption{The response coefficients (a) $|S_3(\omega)|$, (b)
    $|S_4(\omega)|$, and (c) $|S_5(\omega)|$  for relaxation parameters $\Gamma=0.5$~meV and
    $33$~meV  at temperatures $T=0$ and $300$~K and chemical potential
    $\mu=0.3$~eV.  The black dashed curves are the intraband
    contributions from Eq.~(\ref{eq:intra}).}
  \label{fig:pd}
\end{figure}
For the single-mode incident field in Eq.~(\ref{eq:Esgmode}), besides
the SHG, the other second order current is a dc one
\begin{equation}
  \bm J_{\text{dc}}(\bm r,z;t) = \delta(z)
  {\bm q}_0 \left\{S_3(\omega_0) |E_0^\parallel|^2 +
    S_4(\omega_0)|E_0^\perp|^2 \right\} + \delta(z) \hat{\bm z}\times{\bm q}_0 \text{Re} \left[S_5(\omega_0) E_0^\perp
  (E_0^\parallel)^\ast \right] \,,
\end{equation}
where
\begin{equation}
  S_3(\omega_0) =
  4\text{Re}[S_{\mu,T}^{xxxx}(\omega_0,-\omega_0)]\,,\quad
  S_4(\omega_0) =
  4\text{Re}[S^{xyyx}_{\mu,T}(\omega_0,-\omega_0)]\,,\quad
  S_5(\omega_0) =4[S^{xxyy}_{\mu,T}(\omega_0,-\omega_0)-S_{\mu,T}^{xyxy}(-\omega_0,\omega_0)]\,.\label{eq:jdc}
\end{equation}
For these coefficients, the poles exist at $\hbar\omega=0$ or
$\hbar\omega=2|\mu|$. Depending on the electric field polarization,
the dc current can be in the direction of either $\bm q_0$ or both $\bm q_0$ and
$\hat{\bm z}\times\bm q_0$. The latter can only exist when the
electric field has nonzero components along both $\bm q_0$ and $\hat{\bm z}\times\bm
q_0$. We check the limit $\Gamma\to0$ at zero temperature. The
terms in Eq.~(\ref{eq:jdc}) can be approximated as 
\begin{align}
  S_3(\omega) \approx& \text{sgn}(\mu) \sigma_2
  \frac{16[(\hbar\omega)^2-6\mu^2]}{(\hbar\omega)|(\hbar\omega+i\Gamma)^2-4\mu^2|^2}\,,\quad
  S_4(\omega) \approx \text{sgn}(\mu) \sigma_2\frac{-96\mu^2}{(\hbar\omega)|(\hbar\omega+i\Gamma)^2-4\mu^2|^2}\,,\\
  S_5(\omega) \approx& \text{sgn}(\mu) \sigma_2\left\{\frac{1}{\Gamma}\frac{-8i }{(\hbar\omega)^2} + \frac{32[3(\hbar\omega)^2-4\mu^2]\mu^2}{(\hbar\omega)^3|(\hbar\omega+i\Gamma)^2-4\mu^2|^2}\right\}\,.
\end{align}
At $\hbar\omega=\pm2\mu$
the coefficients $S_3(\omega)$ and $S_4(\omega)$ diverge as $\Gamma^{-2}$ for small enough $\Gamma$.
So for sufficiently small $\Gamma$ the small $q$ expansion applied
to Eq. (\ref{eq:2ndv}) becomes suspicious, and the $\bm q$ dependence in
the denominator of that equation should be considered explicitly. For
small frequencies, both coefficients also diverge as
$(\hbar\omega)^{-1}$.  These divergences are associated with the resonant
photon drag effect, as discussed by Entin {\it et al.} \cite{Phys.Rev.B_81_165441_2010_Entin}. For
$S_{5}(\omega)$ there is an additional $\Gamma^{-1}$ divergence that
arises for all photon energy. It only contributes to $J_{dc}$ when
$E_{0}^{\perp}$ and $E_{0}^{\parallel}$ have different phases, which
requires elliptically polarized light. This divergence shows that the dc induced
current described by $S_{5}(\omega)$ behaves as a one-color injected
current, similar to that observed in semiconductors without inversion
symmetry \cite{CoherentControl_Driel}; here the interference between the two transition
amplitudes that can lead to an injected current is associated with
the $E_{0}^{\perp}$ and $E_{0}^{\parallel}$ components of the electric
field.

In Fig.~\ref{fig:pd} we show the response coefficients
$|S_3(\omega)|$, $|S_4(\omega)|$, and $|S_5(\omega)|$ for relaxation parameters $\Gamma=0.5$ and
$33$~meV at temperatures $T=0$ and $300$~K, and chemical
potential $\mu=0.3$~eV. The peaks appearing at $\hbar\omega=0$ and
$\hbar\omega=2|\mu|$ are obvious. Similar to the behavior of
$S_2(\omega)$ in Fig.~\ref{fig:shg} (b),
$|S_4(\omega)|$ in Fig.~\ref{fig:pd} (b) also shows a dip at $\hbar\omega=|\mu|$ at zero
temperature. At finite temperature, the frequency of that dip changes. At zero
temperature, $S_3(\omega)$ and $S_4(\omega)$ show a very weak 
dependence on the relaxation parameters for $\hbar\omega$ away from
the resonances, while $S_5(\omega)$ shows a significant dependence, and
indicates the injection process. The effect of increasing temperature
on $S_3(\omega)$ and $S_4(\omega)$ is significant
for most of the frequencies studied. 

\begin{figure}[!htpb]
  \centering
  \includegraphics[width=15.9cm]{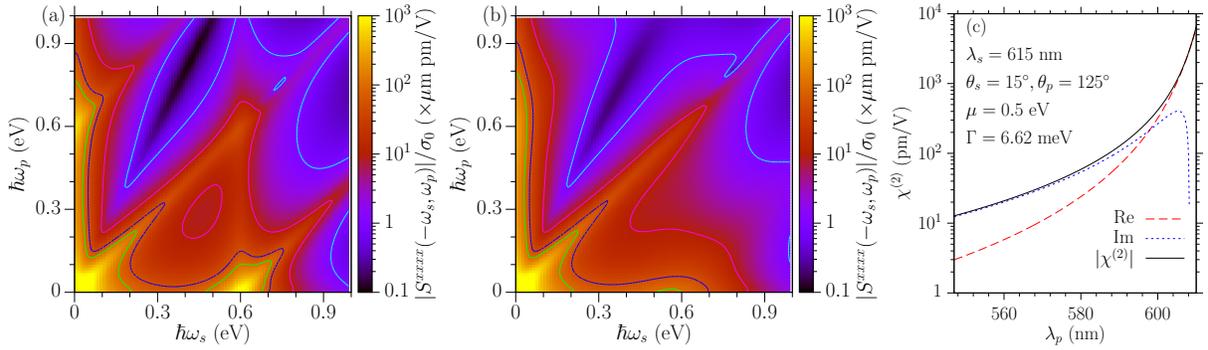}  
  \caption{The contour plot of the response coefficients
    $|S^{xxxx}(-\omega_s,\omega_p)|$ at (a) $T=0$~K and (b)
    $T=300$~K for $\mu=0.3$~eV and $\Gamma=33$~meV. The contour
    lines correspond to the values $1$, $10$, $50$, and $100$ in the
    units indicated. (c) An effective $\chi^{(2)}(-\omega_s,-q_s\hat{\bm
      x};\omega_p,q_p\hat{\bm x})$ with the parameters taken from the
    experiment\cite{Nat.Phys._12_124_2015_Constant} by Constant {\it et al.}. Here
    $\omega_{i}={2\pi c}/{\lambda_{i}}$, and
    $q_{i}=\omega_{i}/c \cos\theta_{i}$ with $i=s,p$. The other
    parameters are $\mu=0.5$~eV, $\Gamma=6.62$~meV, and $T=0$~K.
  }
  \label{fig:dfg}
\end{figure}

\subsection*{Difference frequency generation}
In the presence of a strong pump field $\bm E_{\bm q_p\omega_p}$ at
$\bm q_p$ and $\omega_p$, the injected signal field
$E_{\bm q_s(-\omega_s)}$ at $\bm q_s$ and $-\omega_s$ can lead to light emitted at  $(\bm q_p+\bm
q_s, \omega_p-\omega_s)$ from the second order nonlinear process. In Fig.~\ref{fig:dfg} we show the dependence of
$|S_A^{xxxx}(-\omega_s,\omega_p)|$ on $\hbar\omega_s$ and
$\hbar\omega_p$ at $T=0$ and $T=300$~K for  $\mu=0.3$~eV and
$\Gamma=33$~meV. At zero temperature, large values are
observed around any of $|\hbar\omega_s-\hbar\omega_p|=0$ or $2|\mu|$,
corresponding to the possible poles. Around the line
$\hbar\omega_s=\hbar\omega_p$, the response is rather large due to the
small difference frequency. It has been proposed that this large
signal could be used to  excite 
THz plasmons in graphene \cite{Phys.Rev.Lett._112_55501_2014_Yao}, an
effect reported in an experimental study
\cite{Nat.Phys._12_124_2015_Constant}. 

When exciting of layer structures, the in-plane wave vector can change with the
incident angle while keeping the incident frequency fixed; thus it is possible to find
parameters which satisfy $|\omega_p-\omega_s|/|\bm
q_p+\bm q_s|<c$ as $\omega_p$ and $\omega_s$ get close. The frequency of the emitted
light can then match the plasmon
resonance, which is determined by the linear
conductivity, and the emitted signal can be greatly
enhanced \cite{Nat.Phys._12_124_2015_Constant}. Furthermore, around the condition $\omega_p-\omega_s \sim v_F|\bm
q_p+\bm q_s|$, the second order response can also show a strong $\bm
q$ dependence, where the expansion of the conductivity as Taylor
series of $\bm q$ may not be appropriate.  With finite
temperature and finite relaxation parameters, such dependence can be
further blurred and broadened, which may make expansion
possible. 

We estimate the
  effective susceptibility $\chi^{(2)}(-\omega_s,-q_s\hat{\bm
      x};\omega_p,q_p\hat{\bm x})=\sigma^{(2);xxx}(-\omega_s,-q_s\hat{\bm
      x};\omega_p,q_p\hat{\bm
      x}) / (-i(\omega_p-\omega_s)\epsilon_0d_{\text{gr}})$ with $\omega_i=2\pi c/\lambda_i$,
  $q_i=\omega_i/c \cos\theta_i$ for $i=s,p$, which may be related to
  the experiment \cite{Nat.Phys._12_124_2015_Constant} by Constant
  {\it et al.}. The parameters from the
  experiment are taken as  $\mu=0.5$~eV, $\Gamma=6.62$~meV,
  $\lambda_s=615$~nm, $\theta_s=125^\circ$, and
  $\theta_p=15^\circ$. Note that the results at room temperature are
almost the same as those at zero temperature, which are shown in 
  Fig.~\ref{fig:dfg}(c). Our calculated values
are orders of magnitude smaller than the value extracted from the experiment, which
is about $10^5$~pm/V for
their resonant wavelength $\lambda_p\sim586$~nm. For our parameters,
$\omega_p-\omega_s\ge v_F(q_p-q_s)$ is valid as $\lambda_p\le
612$~nm. At $\lambda_p=586$~nm, $\omega_p-\omega_s$ is about 10 times
larger than $v_F(q_p-q_s)$, which means our approximation should work
at this wavelength. The reason for the discrepancy between the
calculation and the experimental results is not yet clear, and its
clarification probably requires both a more detailed analysis of the
experiment and a theory beyond the single particle approximation. 

\section*{Discussion}
We have separated the contributions of the magnetic dipole-like and
electric quadrupole-like effects to the second order nonlinearities
of monolayer graphene. Using the linear dispersion approximations, we
obtained analytic expressions for the second order conductivities, which
show strong dependence on chemical potential and temperature. We 
quantitatively analyze the predictions for different second order phenomena, including
second harmonic generation, one photon dc current generation, and
difference frequency generation.  Although these effects, forbidden at
the level of the electric dipole approximation, are intrinsically weak, the predicted second order
responses of graphene are very strong, with effective response
coefficients much larger than those for many materials where the
electric dipole effects are allowed.  At low
temperature and with weak relaxation, the calculated second harmonic generation coefficients can
be as large as $10^5$~pm/V for a resonant response at
$\hbar\omega=2|\mu|=0.6$~eV. However, this value decreases to the
order of magnitude of $10^2$~pm/V at room temperature, or in the
presence of strong relaxation. The strength of the second order
response coefficients can be effectively tuned by
applying a gate voltage to graphene for tuning its chemical potential, a strategy which may be used to design
 photonic devices with new functionalities. Finally, we mention that
 the third order nonlinearities, as calculated ealier \cite{Phys.Rev.B_91_235320_2015_Cheng} from
the kind of independent particle approximation applied here, are approximately two orders of
magnitude smaller that those reported in experimental studies. Thus,
it may be that the forbidden second order response in graphene is even larger than the predictions in this
work indicated, and further experimental studies would certainly be in order. 

\section*{Methods}
\subsection*{Two-band tight binding model}
The widely used two-band tight binding model is based on the carbon $2p_z$ orbitals $\varphi(\bm r,z)$, in which the
eigen states and eigen energies of $H_0$ can be written as
\begin{equation}
  \psi_{s\bm
    k}(\bm r,z)=\sum_{\alpha=A,B}c_{s\bm k}^\alpha
  \Phi_{\alpha\bm k}(\bm r,z)\,,\quad  \text{ with }
  c_{s\bm k} =\begin{pmatrix}c_{s\bm k}^A\\c_{s\bm k}^B
  \end{pmatrix}
  =\frac{1}{\sqrt{2}}\begin{pmatrix} s\frac{f_{\bm k}}{|f_{\bm k}|} \\ 1
  \end{pmatrix}\,,
\end{equation}
where $s=\pm$ is the band index, $\bm k=k_x\hat{\bm x}+k_y\hat{\bm y}$
is the two dimensional wave vector, $f_{\bm
  k}=1+e^{-i\bm k\cdot\bm a_1}+e^{-i\bm k\cdot\bm a_2}$ is the
structure factor with primitive lattice vectors $\bm
a_1=(\sqrt{3}\hat{\bm x}-\hat{\bm y})a_0/2$ and $\bm
a_2=(\sqrt{3}\hat{\bm x}+\hat{\bm y})a_0/2$, 
$a_0=2.46$\AA~is the lattice constant, $\gamma_0=2.7$~eV is the hopping energy between nearest
neighbours, and $\Phi_{\alpha\bm k}(\bm
r,z)=\frac{\sqrt{\Omega}}{2\pi}e^{i\bm k\cdot\bm r}\sum_{nm}e^{i\bm
  k\cdot \bm R_{nm}}\varphi(\bm r-\bm R_{nm}-\bm\tau_\alpha,z)$ with $\bm
R_{nm}=n\bm a_1+m\bm a_2$, $\bm\tau_A=0$, and $\bm\tau_B=(\bm a_1+\bm
a_2)/3$. Note that $\varphi(\bm r,z)$ is a function well localized around
$z=0$. The eigen energies are 
\begin{equation}
  \varepsilon_{s\bm k}= s\gamma_0|f_{\bm k}|\,,
\end{equation}
From the localized nature of $\varphi(\bm r,z)$ it follows that the matrix elements of a plane wave are
\begin{equation}
  D_{s_1\bm k_1; s_2\bm k_2} = \int d\bm r \psi_{s_1\bm k_1}^\ast(\bm
  r)e^{i(\bm k_1-\bm k_2)\cdot\bm r}\psi_{s_2\bm k_2}(\bm r) = \frac{1}{2}\left[s_1s_2\frac{f_{\bm
      k_1}^\ast f_{\bm k_2}}{|f_{\bm k_1}f_{\bm k_2}|} + e^{-i(\bm
    k_2-\bm k_1)\cdot\bm\tau_B}\right]\,,
\end{equation}
Under the linear dispersion approximations around the Dirac points, the current
density operator can be defined as
\begin{equation}
\hat{\bm J}_{\bm q} = \frac{e}{2} \left( \hat{\bm v} e^{-i\bm
    q\cdot\bm r} + e^{-i\bm q\cdot\bm r}\hat{\bm v}\right)\,,
\end{equation}
with matrix elements 
\begin{equation}
  \int d\bm r  \psi_{s_1\bm k_1}^\ast(\bm
  r) \hat{J}^d_{\bm q}\psi_{s_2\bm k_2}(\bm r) = e {\cal V}^d_{s_1\bm
    k_1;s_2\bm k_1+\bm q} \delta(\bm k_2-\bm k_1-\bm q)\,.
\end{equation}

\subsection*{Ward identity}
Since the gauge potentials $\phi(\bm r,t;z) = -\partial_t g(\bm
r,t;z)$ and $\bm A(\bm r,t;z)=\bm\nabla_{\bm r} g(\bm r,t;z) $ yields 
zero physical electromagnetic field, they will not induce any changes of the
system. Thus substituting these potentials in Eq.~(\ref{eq:defw2}) 
leads to the Ward identity
\begin{equation}
  W^{(2);d\alpha 0}(\bm q_1,\omega_1;\bm q_2,\omega_2) \omega_2 + W^{(2);d\alpha b}(\bm q_1,\omega_1;\bm q_2,\omega_2) q_2^b=0\,.\label{eq:ward}
\end{equation}
Without loss of the generality, we substitute $W^{(2);d\alpha0} =
-W^{(2);d\alpha b}q_2^b/\omega_2$ into Eq.~(\ref{eq:defw2}); after
appropriate rearrangement, the dependence on electric field $E^b_{\bm
  q\omega}=-iq^b{\cal A}^0_{\bm q\omega}+i\omega{\cal A}^b_{\bm
  q\omega}$ can be extracted. With a similar derivation for ${\cal
  A}^\alpha_{\bm q\omega}$ and then comparing with the expansion in
Eq.~(\ref{eq:defsigma2}), we find
\begin{equation}
  \sigma^{(2);dab}(\bm q_1,\omega_1;\bm q_2,\omega_2) =
  \frac{1}{i\omega_1 i \omega_2} W^{(2);dab}(\bm q_1,\omega_2;\bm q_2,\omega_2)\,.\label{eq:md2}
\end{equation}
Then using the Ward identity we also have
\begin{equation}
  W^{(2);d00}(\bm q_1,\omega_1;\bm q_2,\omega_2) = (-i)^2q_1^aq_2^b
  \sigma^{(2);dab}(\bm q_1,\omega_1;\bm q_2,\omega_2)\,.\label{eq:md1}
\end{equation}

%\bibliography{/home/cheng/REFERENCE/BIB/BIB}

\begin{thebibliography}{10}
\expandafter\ifx\csname url\endcsname\relax
  \def\url#1{\texttt{#1}}\fi
\expandafter\ifx\csname urlprefix\endcsname\relax\def\urlprefix{URL }\fi
\providecommand{\bibinfo}[2]{#2}
\providecommand{\eprint}[2][]{\url{#2}}

\bibitem{Nat.Photon._4_611_2010_Bonaccorso}
\bibinfo{author}{Bonaccorso, F.}, \bibinfo{author}{Sun, Z.},
  \bibinfo{author}{Hasan, T.} \& \bibinfo{author}{Ferrari, A.~C.}
\newblock \bibinfo{title}{Graphene photonics and optoelectronics}.
\newblock \emph{\bibinfo{journal}{Nat. Photon.}} \textbf{\bibinfo{volume}{4}},
  \bibinfo{pages}{611--622} (\bibinfo{year}{2010}).
%\newblock \urlprefix\url{http://dx.doi.org/10.1038/nphoton.2010.186}.

\bibitem{J.Mater.Chem._21_3335_2011_Liu}
\bibinfo{author}{Liu, H.}, \bibinfo{author}{Liu, Y.} \& \bibinfo{author}{Zhu,
  D.}
\newblock \bibinfo{title}{Chemical doping of graphene}.
\newblock \emph{\bibinfo{journal}{J. Mater. Chem.}}
  \textbf{\bibinfo{volume}{21}}, \bibinfo{pages}{3335--3345}
  (\bibinfo{year}{2011}).
%\newblock \urlprefix\url{http://dx.doi.org/10.1039/c0jm02922j}.

\bibitem{Science_320_206_2008_Wang}
\bibinfo{author}{Wang, F.} \emph{et~al.}
\newblock \bibinfo{title}{Gate-variable optical transitions in graphene}.
\newblock \emph{\bibinfo{journal}{Science}} \textbf{\bibinfo{volume}{320}},
  \bibinfo{pages}{206--209} (\bibinfo{year}{2008}).

\bibitem{Sci.Rep._4_6559_2014_Zhang}
\bibinfo{author}{Zhang, Q.} \emph{et~al.}
\newblock \bibinfo{title}{Graphene surface plasmons at the near-infrared
  optical regime}.
\newblock \emph{\bibinfo{journal}{Sci. Rep.}} \textbf{\bibinfo{volume}{4}},
  \bibinfo{pages}{6559} (\bibinfo{year}{2014}).
%\newblock \urlprefix\url{http://dx.doi.org/10.1038/srep06559}.

\bibitem{Europhys.Lett._79_27002_2007_Mikhailov}
\bibinfo{author}{Mikhailov, S.~A.}
\newblock \bibinfo{title}{Non-linear electromagnetic response of graphene}.
\newblock \emph{\bibinfo{journal}{Europhys. Lett.}}
  \textbf{\bibinfo{volume}{79}}, \bibinfo{pages}{27002} (\bibinfo{year}{2007}).
%\newblock \urlprefix\url{http://stacks.iop.org/0295-5075/79/i=2/a=27002}.

\bibitem{NewJ.Phys._16_53014_2014_Cheng}
\bibinfo{author}{Cheng, J.~L.}, \bibinfo{author}{Vermeulen, N.} \&
  \bibinfo{author}{Sipe, J.~E.}
\newblock \bibinfo{title}{Third order optical nonlinearity of graphene}.
\newblock \emph{\bibinfo{journal}{New J. Phys.}} \textbf{\bibinfo{volume}{16}},
  \bibinfo{pages}{053014} (\bibinfo{year}{2014})%.
%\newblock \urlprefix\url{http://dx.doi.org/10.1088/1367-2630/16/5/053014}.
\newblock \bibinfo{note}{{}; New J. Phys. {\bf 18}, 029501 (2016).}

\bibitem{Phys.Rev.B_91_235320_2015_Cheng}
\bibinfo{author}{Cheng, J.~L.}, \bibinfo{author}{Vermeulen, N.} \&
  \bibinfo{author}{Sipe, J.~E.}
\newblock \bibinfo{title}{Third-order nonlinearity of graphene: Effects of
  phenomenological relaxation and finite temperature}.
\newblock \emph{\bibinfo{journal}{Phys. Rev. B}} \textbf{\bibinfo{volume}{91}},
  \bibinfo{pages}{235320} (\bibinfo{year}{2015})%.
\newblock \bibinfo{note}{{}; Phys. Rev. B {\bf 93}, 039904 (2016).}

\bibitem{Phys.Rev.B_93_85403_2016_Mikhailov}
\bibinfo{author}{Mikhailov, S.~A.}
\newblock \bibinfo{title}{Quantum theory of the third-order nonlinear
  electrodynamic effects of graphene}.
\newblock \emph{\bibinfo{journal}{Phys. Rev. B}} \textbf{\bibinfo{volume}{93}},
  \bibinfo{pages}{085403} (\bibinfo{year}{2016}).
%\newblock \urlprefix\url{http://link.aps.org/doi/10.1103/PhysRevB.93.085403}.

\bibitem{Phys.Rep._535_101_2014_Glazov}
\bibinfo{author}{Glazov, M.} \& \bibinfo{author}{Ganichev, S.}
\newblock \bibinfo{title}{High frequency electric field induced nonlinear
  effects in graphene}.
\newblock \emph{\bibinfo{journal}{Phys. Rep.}} \textbf{\bibinfo{volume}{535}},
  \bibinfo{pages}{101--138} (\bibinfo{year}{2014}).
%\newblock \urlprefix\url{http://dx.doi.org/10.1016/j.physrep.2013.10.003}.

\bibitem{Opt.Express_22_15868_2014_Cheng}
\bibinfo{author}{Cheng, J.~L.}, \bibinfo{author}{Vermeulen, N.} \&
  \bibinfo{author}{Sipe, J.~E.}
\newblock \bibinfo{title}{Dc current induced second order optical nonlinearity
  in graphene}.
\newblock \emph{\bibinfo{journal}{Opt. Express}} \textbf{\bibinfo{volume}{22}},
  \bibinfo{pages}{15868--15876} (\bibinfo{year}{2014}).
%\newblock
%  \urlprefix\url{http://www.opticsexpress.org/abstract.cfm?URI=oe-22-13-15868}.

\bibitem{J.Phys.Condens.Matter_20_384204_Mikhailov}
\bibinfo{author}{Mikhailov, S.~A.} \& \bibinfo{author}{Ziegler, K.}
\newblock \bibinfo{title}{Nonlinear electromagnetic response of graphene:
  frequency multiplication and the self-consistent-field effects}.
\newblock \emph{\bibinfo{journal}{J. Phys. Condens. Matter}}
  \textbf{\bibinfo{volume}{20}}, \bibinfo{pages}{384204}
  (\bibinfo{year}{2008}).
%\newblock \urlprefix\url{http://stacks.iop.org/0953-8984/20/i=38/a=384204}.

\bibitem{JETPLett._93_366_2011_Glazov}
\bibinfo{author}{Glazov, M.}
\newblock \bibinfo{title}{Second harmonic generation in graphene}.
\newblock \emph{\bibinfo{journal}{JETP Lett.}} \textbf{\bibinfo{volume}{93}},
  \bibinfo{pages}{366--371} (\bibinfo{year}{2011}).
%\newblock \urlprefix\url{http://dx.doi.org/10.1134/S0021364011070046}.

\bibitem{Phys.Rev.B_84_45432_2011_Mikhailov}
\bibinfo{author}{Mikhailov, S.~A.}
\newblock \bibinfo{title}{Theory of the giant plasmon-enhanced second-harmonic
  generation in graphene and semiconductor two-dimensional electron systems}.
\newblock \emph{\bibinfo{journal}{Phys. Rev. B}} \textbf{\bibinfo{volume}{84}},
  \bibinfo{pages}{045432} (\bibinfo{year}{2011}).
%\newblock \urlprefix\url{http://dx.doi.org/10.1103/PhysRevB.84.045432}.

\bibitem{Appl.Phys.Lett._95_261910_2009_Dean}
\bibinfo{author}{Dean, J.~J.} \& \bibinfo{author}{van Driel, H.~M.}
\newblock \bibinfo{title}{Second harmonic generation from graphene and
  graphitic films}.
\newblock \emph{\bibinfo{journal}{Appl. Phys. Lett.}}
  \textbf{\bibinfo{volume}{95}}, \bibinfo{pages}{261910}
  (\bibinfo{year}{2009}).
%\newblock \urlprefix\url{http://link.aip.org/link/?APL/95/261910/1}.

\bibitem{Phys.Rev.B_82_125411_2010_Dean}
\bibinfo{author}{Dean, J.~J.} \& \bibinfo{author}{van Driel, H.~M.}
\newblock \bibinfo{title}{Graphene and few-layer graphite probed by
  second-harmonic generation: Theory and experiment}.
\newblock \emph{\bibinfo{journal}{Phys. Rev. B}} \textbf{\bibinfo{volume}{82}},
  \bibinfo{pages}{125411} (\bibinfo{year}{2010}).
%\newblock \urlprefix\url{http://dx.doi.org/10.1103/PhysRevB.82.125411}.

\bibitem{Phys.Rev.B_85_121413_2012_Bykov}
\bibinfo{author}{Bykov, A.~Y.}, \bibinfo{author}{Murzina, T.~V.},
  \bibinfo{author}{Rybin, M.~G.} \& \bibinfo{author}{Obraztsova, E.~D.}
\newblock \bibinfo{title}{Second harmonic generation in multilayer graphene
  induced by direct electric current}.
\newblock \emph{\bibinfo{journal}{Phys. Rev. B}} \textbf{\bibinfo{volume}{85}},
  \bibinfo{pages}{121413} (\bibinfo{year}{2012}).
%\newblock \urlprefix\url{http://link.aps.org/doi/10.1103/PhysRevB.85.121413}.

\bibitem{NanoLett._13_2104_2013_An}
\bibinfo{author}{An, Y.~Q.}, \bibinfo{author}{Nelson, F.},
  \bibinfo{author}{Lee, J.~U.} \& \bibinfo{author}{Diebold, A.~C.}
\newblock \bibinfo{title}{Enhanced optical second-harmonic generation from the
  current-biased graphene/sio 2 /si(001) structure}.
\newblock \emph{\bibinfo{journal}{Nano Lett.}} \textbf{\bibinfo{volume}{13}},
  \bibinfo{pages}{2104--2109} (\bibinfo{year}{2013}).
%\newblock \urlprefix\url{http://dx.doi.org/10.1021/nl4004514}.

\bibitem{Phys.Rev.B_89_115310_2014_An}
\bibinfo{author}{An, Y.~Q.}, \bibinfo{author}{Rowe, J.~E.},
  \bibinfo{author}{Dougherty, D.~B.}, \bibinfo{author}{Lee, J.~U.} \&
  \bibinfo{author}{Diebold, A.~C.}
\newblock \bibinfo{title}{Optical second-harmonic generation induced by
  electric current in graphene on si and {SiC} substrates}.
\newblock \emph{\bibinfo{journal}{Phys. Rev. B}} \textbf{\bibinfo{volume}{89}},
  \bibinfo{pages}{115310} (\bibinfo{year}{2014}).
%\newblock \urlprefix\url{http://dx.doi.org/10.1103/PhysRevB.89.115310}.

\bibitem{Appl.Phys.Lett._105_151605_2014_Lin}
\bibinfo{author}{Lin, K.-H.}, \bibinfo{author}{Weng, S.-W.},
  \bibinfo{author}{Lyu, P.-W.}, \bibinfo{author}{Tsai, T.-R.} \&
  \bibinfo{author}{Su, W.-B.}
\newblock \bibinfo{title}{Observation of optical second harmonic generation
  from suspended single-layer and bi-layer graphene}.
\newblock \emph{\bibinfo{journal}{Appl. Phys. Lett.}}
  \textbf{\bibinfo{volume}{105}}, \bibinfo{pages}{151605}
  (\bibinfo{year}{2014}).
%\newblock \urlprefix\url{http://dx.doi.org/10.1063/1.4898065}.

\bibitem{NanoLett._12_2032_2012_Wu}
\bibinfo{author}{Wu, S.} \emph{et~al.}
\newblock \bibinfo{title}{Quantum-enhanced tunable second-order optical
  nonlinearity in bilayer graphene}.
\newblock \emph{\bibinfo{journal}{Nano Lett.}} \textbf{\bibinfo{volume}{12}},
  \bibinfo{pages}{2032--2036} (\bibinfo{year}{2012}).
%\newblock \urlprefix\url{http://pubs.acs.org/doi/abs/10.1021/nl300084j}.

\bibitem{J.Nanophoton._6_61702_2012_Avetissian}
\bibinfo{author}{Avetissian, H.~K.}, \bibinfo{author}{Avetissian, A.~K.},
  \bibinfo{author}{Mkrtchian, G.~F.} \& \bibinfo{author}{Sedrakian, K.~V.}
\newblock \bibinfo{title}{Multiphoton resonant excitation of fermi-dirac sea in
  graphene at the interaction with strong laser fields}.
\newblock \emph{\bibinfo{journal}{J. Nanophoton.}}
  \textbf{\bibinfo{volume}{6}}, \bibinfo{pages}{061702} (\bibinfo{year}{2012}).
%\newblock \urlprefix\url{http://dx.doi.org/10.1117/1.JNP.6.061702}.

\bibitem{Ganichev_2010}
\bibinfo{author}{Ganichev, S.~D.} \emph{et~al.}
\newblock \bibinfo{title}{Photon helicity driven currents in graphene}.
\newblock \emph{\bibinfo{journal}{35th International Conference on Infrared,
  Millimeter, and Terahertz Waves}}  (\bibinfo{year}{2010}).
%\newblock \urlprefix\url{http://dx.doi.org/10.1109/ICIMW.2010.5612541}.

\bibitem{Phys.Rev.B_81_165441_2010_Entin}
\bibinfo{author}{Entin, M.~V.}, \bibinfo{author}{Magarill, L.~I.} \&
  \bibinfo{author}{Shepelyansky, D.~L.}
\newblock \bibinfo{title}{Theory of resonant photon drag in monolayer
  graphene}.
\newblock \emph{\bibinfo{journal}{Phys. Rev. B}} \textbf{\bibinfo{volume}{81}},
  \bibinfo{pages}{165441} (\bibinfo{year}{2010}).
%\newblock \urlprefix\url{http://dx.doi.org/10.1103/PhysRevB.81.165441}.

\bibitem{Phys.Rev.Lett._112_55501_2014_Yao}
\bibinfo{author}{Yao, X.}, \bibinfo{author}{Tokman, M.} \&
  \bibinfo{author}{Belyanin, A.}
\newblock \bibinfo{title}{Efficient nonlinear generation of thz plasmons in
  graphene and topological insulators}.
\newblock \emph{\bibinfo{journal}{Phys. Rev. Lett.}}
  \textbf{\bibinfo{volume}{112}}, \bibinfo{pages}{055501}
  (\bibinfo{year}{2014}).
%\newblock
%  \urlprefix\url{http://link.aps.org/doi/10.1103/PhysRevLett.112.055501}.

\bibitem{Nat.Phys._12_124_2015_Constant}
\bibinfo{author}{Constant, T.~J.}, \bibinfo{author}{Hornett, S.~M.},
  \bibinfo{author}{Chang, D.~E.} \& \bibinfo{author}{Hendry, E.}
\newblock \bibinfo{title}{All-optical generation of surface plasmons in
  graphene}.
\newblock \emph{\bibinfo{journal}{Nat. Phys.}} \textbf{\bibinfo{volume}{12}},
  \bibinfo{pages}{124--127} (\bibinfo{year}{2016}).
%\newblock \urlprefix\url{http://dx.doi.org/10.1038/nphys3545}.

\bibitem{Phys.Rev.B_93_235422_2016_Tokman}
\bibinfo{author}{Tokman, M.}, \bibinfo{author}{Wang, Y.},
  \bibinfo{author}{Oladyshkin, I.}, \bibinfo{author}{Kutayiah, A.~R.} \&
  \bibinfo{author}{Belyanin, A.}
\newblock \bibinfo{title}{Laser-driven parametric instability and generation of
  entangled photon-plasmon states in graphene}.
\newblock \emph{\bibinfo{journal}{Phys. Rev. B}} \textbf{\bibinfo{volume}{93}},
  \bibinfo{pages}{235422} (\bibinfo{year}{2016}).
%\newblock \urlprefix\url{http://dx.doi.org/10.1103/PhysRevB.93.235422}.

%\bibitem{noteadd}
%\newblock \bibinfo{journal}{After the submission, we became aware of related works in preprints [\citeonline{Phys.Rev.B_94_195442_2016_Wang,arXiv:1610.04854}]. Overlapping results in these papers are in agreement in the absence of relaxation time. }

\bibitem{Phys.Rev.B_94_195442_2016_Wang}
\bibinfo{author}{Wang, Y.}, \bibinfo{author}{Tokman, M.} \&
  \bibinfo{author}{Belyanin, A.}
\newblock \bibinfo{title}{Second-order nonlinear optical response of graphene}.
\newblock \emph{\bibinfo{journal}{Phys. Rev. B}} \textbf{\bibinfo{volume}{94}},
  \bibinfo{pages}{195442} (\bibinfo{year}{2016}).
%\newblock \urlprefix\url{http://link.aps.org/doi/10.1103/PhysRevB.94.195442}.

\bibitem{arXiv:1610.04854}
\bibinfo{author}{Rostami, H.}, \bibinfo{author}{Katsnelson, M.~I.} \&
  \bibinfo{author}{Polini, M.}
\newblock \bibinfo{title}{Theory of plasmonic effects in nonlinear optics: the
  case of graphene}.
\newblock {\bibinfo{note}{arXiv:1610.04854}}  (\bibinfo{year}{2016}).
%\newblock \urlprefix\url{http://cn.arxiv.org/abs/1610.04854v1}.
%\newblock \eprint{1610.04854}.
  
  
\bibitem{Phys.Rev.B_1_2362_1970_Mermin}
\bibinfo{author}{Mermin, N.}
\newblock \bibinfo{title}{Lindhard dielectric function in the relaxation-time
  approximation}.
\newblock \emph{\bibinfo{journal}{Phys. Rev. B}} \textbf{\bibinfo{volume}{1}},
  \bibinfo{pages}{2362--2363} (\bibinfo{year}{1970}).
%\newblock \urlprefix\url{http://dx.doi.org/10.1103/PhysRevB.1.2362}.

\bibitem{NewJ.Phys._8_318_2006_Wunsch}
\bibinfo{author}{Wunsch, B.}, \bibinfo{author}{Stauber, T.},
  \bibinfo{author}{Sols, F.} \& \bibinfo{author}{Guinea, F.}
\newblock \bibinfo{title}{Dynamical polarization of graphene at finite doping}.
\newblock \emph{\bibinfo{journal}{New J. Phys.}} \textbf{\bibinfo{volume}{8}},
  \bibinfo{pages}{318} (\bibinfo{year}{2006}).
%\newblock \urlprefix\url{http://stacks.iop.org/1367-2630/8/i=12/a=318}.

\bibitem{Phys.Rev.B_75_205418_2007_Hwang}
\bibinfo{author}{Hwang, E.~H.} \& \bibinfo{author}{Das~Sarma, S.}
\newblock \bibinfo{title}{Dielectric function, screening, and plasmons in
  two-dimensional graphene}.
\newblock \emph{\bibinfo{journal}{Phys. Rev. B}} \textbf{\bibinfo{volume}{75}},
  \bibinfo{pages}{205418} (\bibinfo{year}{2007}).
%\newblock \urlprefix\url{http://link.aps.org/doi/10.1103/PhysRevB.75.205418}.

\bibitem{Phys.Rev.B_92_235307_2015_Cheng}
\bibinfo{author}{Cheng, J.~L.}, \bibinfo{author}{Vermeulen, N.} \&
  \bibinfo{author}{Sipe, J.~E.}
\newblock \bibinfo{title}{Numerical study of the optical nonlinearity of doped
  and gapped graphene: From weak to strong field excitation}.
\newblock \emph{\bibinfo{journal}{Phys. Rev. B}} \textbf{\bibinfo{volume}{92}},
  \bibinfo{pages}{235307} (\bibinfo{year}{2015}).
%\newblock \urlprefix\url{http://link.aps.org/doi/10.1103/PhysRevB.92.235307}.

\bibitem{CoherentControl_Driel}
\bibinfo{author}{van Driel, H.~M.} \& \bibinfo{author}{Sipe, J.~E.}
\newblock \bibinfo{title}{Coherent control | applications in semiconductors}.
\newblock In \bibinfo{editor}{Guenther, B.} (ed.)
  \emph{\bibinfo{booktitle}{Encyclopedia of Modern Optics}},
  \bibinfo{pages}{137 -- 143} (\bibinfo{publisher}{Elsevier},
  \bibinfo{address}{Oxford}, \bibinfo{year}{2005}).
%\newblock
%  \urlprefix\url{http://www.sciencedirect.com/science/article/B7GFY-4G82V0V-20/2/a755715a2a5822fa9f7d6d5c04ca3ade}.

%% \bibitem{Phys.Rev.B_94_195442_2016_Wang}
%% \bibinfo{author}{Wang, Y.}, \bibinfo{author}{Tokman, M.} \&
%%   \bibinfo{author}{Belyanin, A.}
%% \newblock \bibinfo{title}{Second-order nonlinear optical response of graphene}.
%% \newblock \emph{\bibinfo{journal}{Phys. Rev. B}} \textbf{\bibinfo{volume}{94}},
%%   \bibinfo{pages}{195442} (\bibinfo{year}{2016}).
%% %\newblock \urlprefix\url{http://link.aps.org/doi/10.1103/PhysRevB.94.195442}.

%% \bibitem{arXiv:1610.04854}
%% \bibinfo{author}{Rostami, H.}, \bibinfo{author}{Katsnelson, M.~I.} \&
%%   \bibinfo{author}{Polini, M.}
%% \newblock \bibinfo{title}{Theory of plasmonic effects in nonlinear optics: the
%%   case of graphene}.
%% \newblock {\bibinfo{note}{arXiv:1610.04854}}  (\bibinfo{year}{2016}).
%% %\newblock \urlprefix\url{http://cn.arxiv.org/abs/1610.04854v1}.
%% %\newblock \eprint{1610.04854}.

\end{thebibliography}
%\input{forbiddensp.bbl}

\section*{Acknowledgements (not compulsory)}
This work has been supported by the EU-FET grant GRAPHENICS (618086),
by the ERC-FP7/2007-2013 grant 336940, by the FWO-Vlaanderen project
G.A002.13N, by the Natural Sciences and Engineering Research Council of
Canada, by VUB-Methusalem, VUB-OZR, and IAP-BELSPO under grant IAP
P7-35.

\section*{Author contributions statement}
J.L.C performed the derivation and calculation, all authors discussed
the idea, analyzed the results, and revised the manuscript. 

\section*{Additional information}
\textbf{Competing financial interests}: The authors declare no competing
financial interests. \\
The present address of J.L.C is : Changchun Institute of Optics, fine Mechanics and
Physics, Chinese Academy of Sciences, 3888 Eastern South Lake Road,
Changchun, Jilin 130033, China. \\
%{\it Note added.} After the submission, we became aware of related works in preprints [\citeonline{Phys.Rev.B_94_195442_2016_Wang,arXiv:1610.04854}]. Overlapping results in these papers are in agreement in the absence of relaxation time. 

\end{document}